\documentclass[%
 reprint,
 amsmath,amssymb,
pra,
]{revtex4-2}
\usepackage{hyperref}
\usepackage{graphicx}% Include figure files
\usepackage{bm}% bold math
\usepackage{float}
\usepackage{amsmath, amssymb}
\usepackage{soul}
\usepackage{physics}
\usepackage{xcolor}
\usepackage{tikz}
\usetikzlibrary{quantikz2}
\renewcommand{\a}{\hat{a}}
\newcommand{\ad}{\hat{a}^\dagger}
\renewcommand{\b}{\hat{b}}
\newcommand{\bd}{\hat{b}^\dagger}
\newcommand\avg[1]{\langle #1\rangle}
\preprint{APS/123-QED}
\begin{document}

\title{Entanglement and coherence dynamics in photonic quantum memristors}% Force line breaks with \\

\author{Alberto Ferrara}
\author{Rosario Lo Franco}%
\affiliation{%
 Dipartimento di Ingegneria, Università degli Studi di Palermo, Viale delle Scienze, 90128 Palermo, Italy
}%

\date{\today}% It is always \today, today,

\begin{abstract}
Memristive systems exhibit dynamics that depend on their past states, making them useful as memory units. 
Recently, quantum memristor models have been proposed and notably, a photonic quantum memristor (PQM) has been experimentally proven. In this work, we explore and characterize various quantum properties that emerge from this specific model of PQM. Firstly, we find that a single PQM displays memristive dynamics on its quantum coherence. Secondly, we analytically show that a network made of two independent PQMs can manifest memory effects on the dynamics of both entanglement and coherence of correlated photons traveling through the network, regardless of their distance, in the hypothesis of negligible external disturbances. Additionally, we build and run a circuit-model of the PQM on a real qubit-based quantum computer (IBM-Q), showing that: (i) this system can effectively be used for non-linear quantum computing under specific conditions, and (ii) digital quantum simulations can reproduce the dynamics of a memristive quantum system in a non-Markovian regime. 
\end{abstract}

\maketitle

%\tableofcontents

\section{Introduction}
The concept of memristor has been introduced in 1971 \cite{chua1971memristor} as a resistor with memory properties. Specifically, a memristor is defined as a two-terminal electrical component, relating magnetic flux linkage $\phi(t)$ to $q(t)$, the amount of charge flowed into the component, through a non-linear relation. This is accompanied by an internal variable, the memristance $M(q(t))$, whose value changes in time and depends exclusively on $q(t)$. As a consequence, the dynamics of the memristor manifests memory, i.e., dynamical regimes that depend on the system's past states. Nowadays, we are used to refer to these systems as non-Markovian \cite{breuer2002theory}.
The interest into memristors notably grew after the first alleged experimental realization \cite{strukov2008missing}, which opened a wide research area focused on characterizing their properties, their potential advantages over classical devices like transistors, and the specifics of physical implementations.
However, doubts over the concept of ideal memristor have been raised \cite{vongehr2015missing} and conclusive proof of their existence might not be found \cite{kim2020experimental}. Nevertheless, the entire class of memristive systems showing memory effects while not being the ideal circuit elements, remains perfectly valid and at the center of a great interest.
In this more general view, a given system which exhibits the typical memristive dynamics can be named memristor \cite{tetzlaff2013memristors}.

Among some recents proposals, quantum memristor models that retain quantum coherence while still manifesting non-linearities in their dynamics \cite{pfeiffer2016quantum} have been studied on different platforms, such as superconducting circuits \cite{salmilehto2017quantum}, trapped-ions \cite{stremoukhov2023proposal}, optical cavities \cite{norambuena2022polariton} and photonic circuits \cite{sanz2018invited}. As a notable result, one specific model of photonic quantum memristor (PQM) has been experimentally proven \cite{spagnolo2022experimental, spagnolo2024quantum}.

In this work, we explore and present some genuine quantum features that emerge when considering two distinct independent photonic quantum memristors (PQMs) in paradigmatic settings. Starting from pre-entangled couples of photons, we discover that the system is capable to exhibit memristive dynamics both on the entanglement between photons and on their quantum coherence. We characterize the memory properties by means of the form factor and show the different dynamics that emerge for various ranges of the internal memory of the memristor. 

As an additional result, we also present a quantum circuit conversion of the PQM, tackling the problem of encoding such bosonic system into a qubit-based quantum architecture.
In general, simulating a physical system by means of a digital quantum simulation is one the most promising applications of quantum computing and among the first to be proposed, dating back to Feynman \cite{feynman2018simulating}. As the size of quantum computers increases, wider possibilities open up,  giving rise to a lively research field \cite{fauseweh2024quantum}. However, it is often found that translating a generic quantum system into a digital quantum system is not a trivial task. 
Here, we execute our code on a real quantum computer \cite{ibm-quantum}, showing that the memristive dynamics can be reproduced through a qubit-based system, making the PQM a good candidate for future physical implementations of quantum computing. 

The paper is organized as follows: 
In Sec.~\ref{sec: system_review} we briefly review some general properties of quantum memristors and outline the PQM \cite{spagnolo2022experimental}, also showing previously unreported dynamical regimes for the quantum coherence.
Sec.~\ref{sec: conversion} is devoted to the digital quantum simulation of the PQM. We shall highlight advantages and limitations of this approach. 
In Sec.~\ref{sec: entangled_section} we present the setting adopted to study the two PQMs assembly and the main results obtained with entangled inputs. We characterize various emerging dynamical regimes for different values of the internal memory period of the memristors. 
Finally, in Sec.~\ref{sec: conclusions} we sum up our findings and provide some prospects on future works.

\section{Single photonic quantum memristor (PQM)}
\label{sec: system_review}
In its general formulation, a memristive system is characterized by the dynamical equations
\begin{equation}
    \begin{split}
        & y (t) = f (s, x, t) x (t), \\
        & \Dot{s} = g (s, x, t),
    \end{split}
    \label{eq: mem_eqsn}
\end{equation}
where $y(t)$ represents the output of our system while $x(t)$ is the input. 
The internal state of the memristor is described by a function $s(t)$, whose derivative, given by $g(s, x, t)$, depends on the input and on the state variable itself. Instead of considering only one degree of freedom, it is possible to add multiple internal state variables to achieve more complex and interesting dynamics \cite{kumar2022dynamical}.

In the original formulation of the memristor \cite{chua1971memristor}, the input is the electric current flowing inside the electrical component, while the output is the voltage at its ends. The internal variable, the memristance, changes in time according to the amount of charge that passed through it in the past (i.e., the integral of $q(t)$ over a fixed time-span). 
This specific input-output relation gives rise to signature hysteresis loops, underlining the history-dependant dynamics of such a system. This means that, given an appropriate time-dependant input, there are two distinct possible values of the output, thus implying additional information on the direction (i.e., past states) of the dynamics. 
Due to the explicit linear dependence of the output on the input, an absence of input results in no output, which gives the hysteresis loop its characteristic pinched appearance.

A first quantum memristive system has been proposed in Ref.~\cite{pfeiffer2016quantum}. Subsequently, a variety of different models on different platforms have been conceived.
The main requirements for a quantum memristor are that it needs to manifest the dynamics described in Eq.~\eqref{eq: mem_eqsn} for the expectation values of some observables and, additionally, it has to coherently process quantum states. 
In these features lies the full potential of the memristor. Specifically, the dynamical equations are non-linear and allow for many applications that require non-Markovian behaviour, as neuromorphic computing. At the same time, coherence must be maintained in order to exploit the advantages of quantum information over classical processes.
This property can be achieved by constructing an open quantum system, with a specific weak system-environment interaction that provides non-linearity while not causing complete decoherence.

The starting point of the present work is the photonic quantum memristor (PQM) originally proposed in Refs.~\cite{sanz2018invited,gonzalez2020quantum} and subsequently experimentally proven in Ref.~\cite{spagnolo2022experimental}. In the following we briefly recall the main features of the PQM. 

\begin{figure}
         \includegraphics[width=0.42\textwidth]{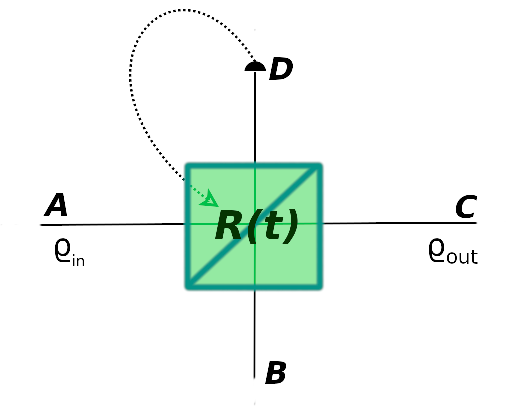}
         \includegraphics[width=0.42\textwidth]{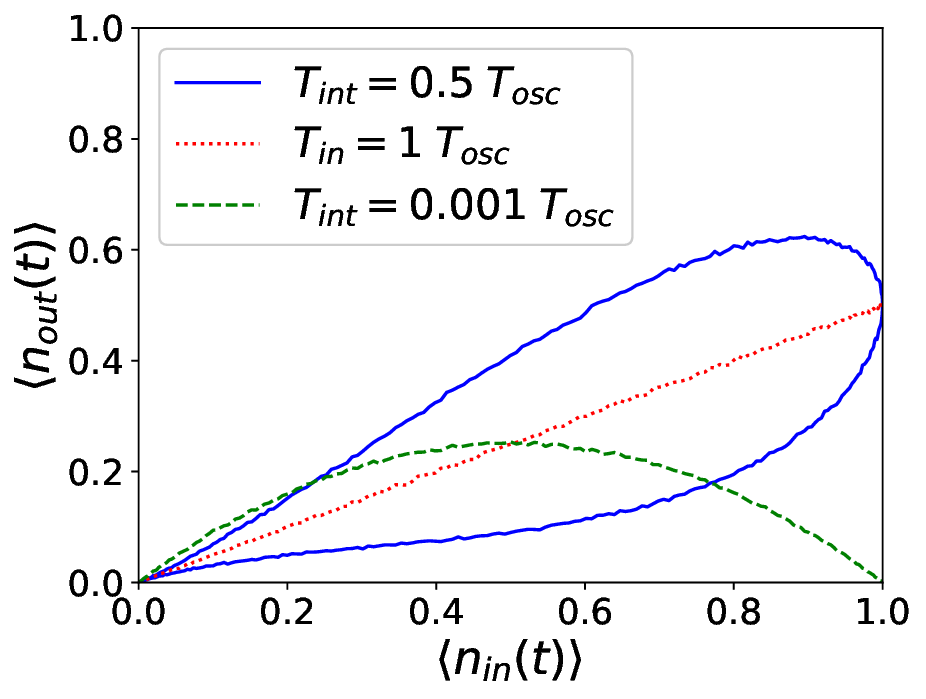}
         \includegraphics[width=0.42\textwidth]{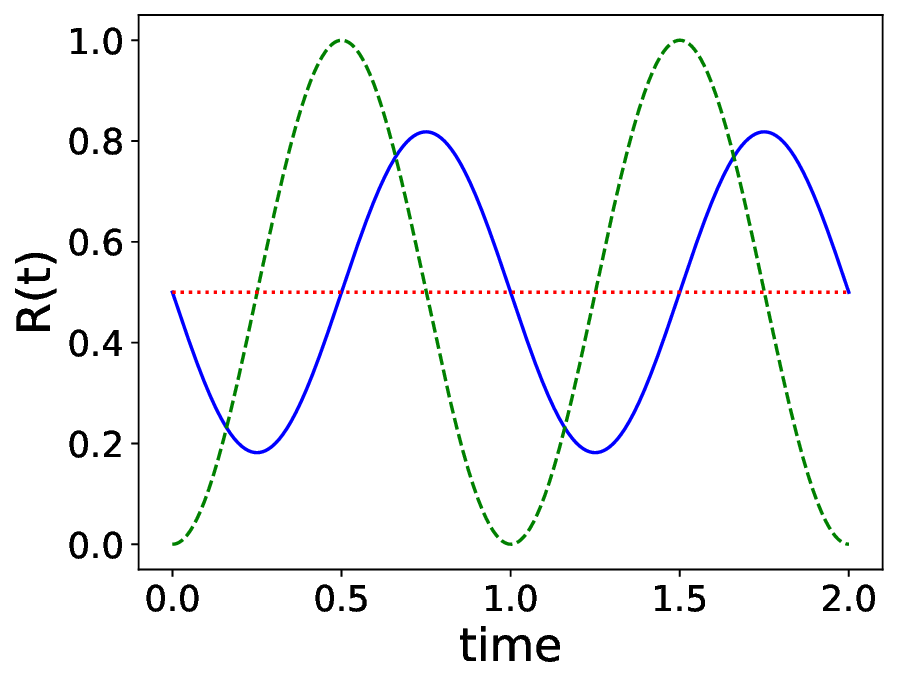}
         \caption{\textbf{Upper panel:} Pictorial representation of the photonic quantum memristor. The studied system is composed by a Beamsplitter with tunable Reflectivity, denoted as $R(t)$. $A$ and $B$ represent the input modes, while $C$ and $D$ are the output modes. An input state $\rho_{in}$ is sent through mode $A$. A measurement is performed on mode $D$ to update the Reflectivity, while the state coming out of mode $C$ represents the output. \textbf{Central panel:} Average output photon number $\avg{{n_\mathrm{out}}(t)}$ vs average input photon number $\avg{{n_\mathrm{in}}(t)}$ at three different ratios of $T_\mathrm{int}/T_\mathrm{osc}$. \textbf{Lower panel:} Reflectivity $R(t)$ as a function of time for an oscillatory input state, as given in Eq.~\eqref{eq: input_state_single}. We can see that for $T_\mathrm{int} = T_\mathrm{osc}$ the reflectivity is constant.. In this scenario, the PQM acts as a $50/50$ beamsplitter. The legend for the central panel applies for the lower panel as well.}
        \label{fig:sumup_single}
\end{figure}

As pictorially displayed in the upper panel of Fig.~\ref{fig:sumup_single}, the system is built around a beamsplitter whose reflectivity $R(t)$ (i.e., the probability that an incident photon is reflected) is tuned over time according to a measurement performed on a part of the system's own output.
The PQM, just like a beamsplitter, takes two distinct inputs, in the form of two optical paths, denoted as $A$ and $B$. Since we are considering a photonic platform, we take a generic time-dependent input state in mode $A$ describing the state of a single photon path, which can be written as a generic two-level system
\begin{equation}
    \ket{\psi_\mathrm{in}} = \alpha (t) \ket{0} + \beta (t) \ket{1}.
    \label{eq: input_state_single}
\end{equation}
In this picture, $\ket{0}$ represents a channel with no photon, while $\ket{1}$ indicates the presence of a single photon in the path. The second one, B, is an ancilla state which we choose as empty in our study. 
This implies that the global input state will be the simple tensor product $\ket{\psi_\mathrm{in}}\otimes\ket{0}$. Being the ancilla state empty, the output at mode $C$ is simply given by the chance that the photon in $\ket{\psi_\mathrm{in}}$ is transmitted. This depends on the reflectivity at some specific time $t$. If we take the average value of the output photon number, we get
\begin{equation}
    \avg{{n_\mathrm{out}}(t)} = (1 - R(t)) \avg{{n_\mathrm{in}}(t)},
    \label{eq: avg_value}
\end{equation}
Here and in the following, we define the average value of the photon number as the expectation value of a given state of the number operator, i.e. $\avg{n_{\psi}} = \bra{\psi} \hat{n} \ket{\psi}$, where $\hat{n} = \sum_{n = 0}^{\infty} n\ket{n}\bra{n}$. We see that Eq.~\eqref{eq: avg_value} has the exact shape of the first memristive relation of Eq.~\eqref{eq: mem_eqsn}. The role of the second relation of Eq.~\eqref{eq: mem_eqsn} is then played by the time evolution of the reflectivity $R(t)$, which is the core of this peculiar system. The time derivative of $R(t)$ is thus fixed as
\begin{equation}\label{DerivativeR}
    \Dot{R}(t) = \avg{{n_\mathrm{in}}(t)} - 0.5 \avg{n_\mathrm{max}}.
\end{equation}
One subtracts half the average maximum value of the photon number so that $\Dot{R}(t)$ can be both positive and negative (notice that, in the single-photon case, $\avg{n_\mathrm{max}}=1$ and $\avg{{n_\mathrm{in}}(t)} = |\beta (t)|^2$). This aspect confers oscillatory behaviour to the reflectivity.
To have reflectivity values well confined between $0$ and $1$, one assumes that our memristor has a memory lasting over a fixed time interval $T_\mathrm{int}$, which represents the characteristic period of the memristor itself. A solution of Eq.~(\ref{DerivativeR}) is given by
\begin{equation}
    R(t) = 0.5 + \frac{1}{T_\mathrm{int}} \int_{t - T_\mathrm{int}}^{t} (\avg{n_\mathrm{in}(t')} - 0.5) dt'.
    \label{eq: reflectivity_equation}
\end{equation}

The memristor gets information about the average value of the input photon number through a measurement made on the ancilla qubit of the system. The output $D$ is used for the feedback of the quantum memristor, whereas the accessible output is provided by the output C (see upper panel of Fig.~\ref{fig:sumup_single}). It can be shown that on mode $D$, the output branch which has not still been used, one obtains
\begin{equation}
    \avg{n_\mathrm{meas}(t)} = R(t) \avg{n_\mathrm{in}(t)} \implies \avg{n_\mathrm{in}(t)} = \frac{\avg{n_\mathrm{meas}(t)}}{R(t)},
    \label{eq: Reflectivity_single}
\end{equation}
which means that by performing a measurement on the non-relevant part of the output, we can estimate the input, thus implementing Eq.~\eqref{eq: reflectivity_equation}. 

To illustrate the memristive property of the system, we can pick a specific choice for the input state of Eq.~(\ref{eq: input_state_single}), such as
\begin{equation}\label{alphabeta}
     \alpha(t) = \cos{\left( \frac{\pi t}{T_\mathrm{osc} } \right)},\quad
     \beta(t) = \sin{\left(\frac{\pi t}{T_\mathrm{osc} }\right)}.
\end{equation}
Explicit integration of Eq.~\eqref{eq: reflectivity_equation} in this setting leads to
\begin{equation}
    \begin{split}
    R(t) = \frac{T_\mathrm{osc}}{T_\mathrm{int}}\dfrac{\sin\left(\frac{2{\pi}\left(t-T_\mathrm{int}\right)}{T_\mathrm{osc}}\right)-\sin\left(\frac{2{\pi}t}{T_\mathrm{osc}}\right)}{4{\pi}}+\dfrac{1}{2},
    \end{split}
\end{equation}
which is plotted, along with the average output photon number $\avg{{n_\mathrm{out}}(t)}$, in Fig.~\ref{fig:sumup_single}.

The density matrix of the system of interest at the output $C$ can be finally obtained by computing the global density matrix of the output (seen as system $C$ + environment $D$) and then tracing out the environment degrees of freedom \cite{spagnolo2022experimental}, that is
\begin{equation}\label{outputC}
        \rho_{\mathrm{out, C}} = 
            \begin{pmatrix}
            \abs{\alpha(t)}^2 + \abs{\beta(t)}^2 R(t)& \alpha^{*} \beta \sqrt{1 - R(t)} \\
            \alpha \beta^{*} \sqrt{1 - R(t)} & \abs{\beta(t)}^2(1 - R(t)) \\
            \end{pmatrix}.
\end{equation}
In the original experimental work \cite{spagnolo2022experimental}, the purity of the output state is shown and compared to the results of the tomography of the quantum state.

\subsection{Coherence dynamics in a PQM}
To further characterize the quantum nature of the single PQM, here we provide the behaviour of quantum coherence during the system evolution, which has not been previously reported. To this aim, we adopt the $l_1$-\textit{norm} of coherence \cite{baumgratz2014quantifying,streltsov2017colloquium}, computed as $\mathcal{C}_{l_{1}}=\sum_{i\neq j}|\rho_{ij}|$. 

For the output and input states of the PQM, given in 
Eqs.~(\ref{outputC}) and (\ref{eq: input_state_single}), respectively, with the specific choices of Eq.~(\ref{alphabeta}), we get
%\begin{equation}
%    \mathcal{C}_{l_{1}}(\rho_{\mathrm{out, C}}) = \sum_{i \neq j} [\rho_{\mathrm{out, C}}]_{ij} = \sqrt{1 - R(t)} (\alpha^{*} \beta + \alpha \beta^{*}).
%\end{equation}
%In the considered example, this leaves us with:
\begin{equation}
    \begin{split}
        & \mathcal{C}_{l_{1}}(\rho_{\mathrm{in, A}}) = 2 \abs{\alpha(t) \beta (t)} = 2 \left|\sin{\left( \frac{t}{T_{\mathrm{osc}} }\pi \right)} \cos{\left( \frac{t}{T_{\mathrm{osc}} }\pi \right)}\right|, \\
        &  \mathcal{C}_{l_{1}}(\rho_{\mathrm{out, C}}) = \mathcal{C}_{l_{1}} (\rho_{\mathrm{in, A}})\sqrt{1 - R(t)}. \\
    \end{split}
    \label{eq: coherence_l1_norm}
\end{equation}
The input coherence, as expected, oscillates in time according to the input state. For our specific choice of probability amplitudes, the coherence oscillates between $0$ (input state with no superposition) and $1$ (input state with equal superposition of $\ket{0}$ and $\ket{1}$). We see that the output coherence oscillates according to the same behaviour, with a multiplicative modulation through a $\sqrt{1 - R(t)}$ factor. This is relevant, because it explicitly shows that even though measurements have been performed on the environment (i.e. output $D$, through which the internal state of the memristive device changes), the relevant part of the output, obtained via partial trace over mode $D$, still maintains quantum coherence. This happens despite that the relevant output in mode $C$ and the environment output in mode $D$ are, in general, entangled when coming out of the memristor. From a practical point of view, this is equivalent to assuming that, for each time step, the measurement is repeated many times until its statistics can be determined. Therefore, the output is a statistical mixture of the possible outcomes in mode C due to the measurement on mode D.

\begin{figure}
        \includegraphics[width=0.48\textwidth]{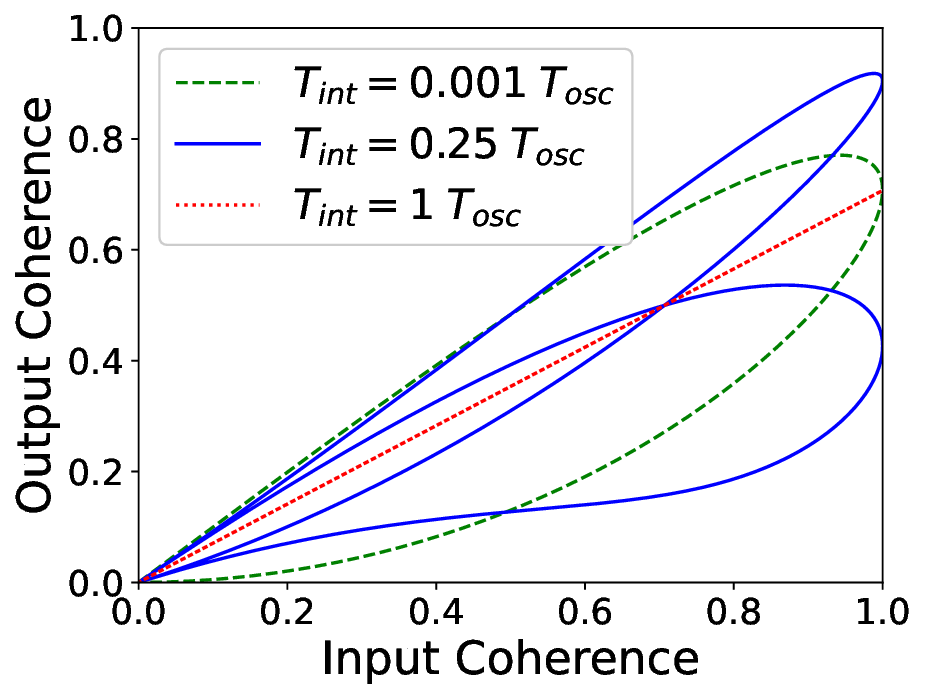}
         \caption{Response curve for the quantum coherence in the single PQM for different values of the integration time $T_\mathrm{int}$. For small periods, we see two hysteresis loops overlapping (green curve). For larger $T_\mathrm{int}$, the two cycles become progressively more separated (blue solid curves), until the high period limit ($T_\mathrm{int}=T_\mathrm{osc}$) when no hysteresis is observed (red dotted line).}
        \label{fig:coherence_single}
\end{figure}

As an additional result, we can easily see in Eqs.~(\ref{eq: coherence_l1_norm}) that the relation between the output coherence and the input coherence has the same shape of the original memristive relation given in Eq.~(\ref{eq: mem_eqsn}). This fact proves that a single PQM is capable of manifesting hysteresis loops also on the dynamics of the single-photon quantum coherence, without making any assumption neither on the evolution of $\alpha(t)$, $\beta(t)$, nor on the specific function adopted to update $R(t)$. Some possible dynamical regimes of PQM coherence are displayed in Fig.~\ref{fig:coherence_single}.

\section{Qiskit circuit conversion of the PQM}
\label{sec: conversion}
In this section we show that the PQM can be used to coherently process qubit states. Firstly, We propose a circuit model to simulate the action of a single PQM over a two-qubit system. Secondly, we show that this can be effectively run on an IBM quantum computer \cite{ibm-quantum}, thus making quantum computing platforms capable of reproducing memory effects via hysteresis loops. We also point out the limitations of this approach, which are due to some specific properties of quantum photonic systems.

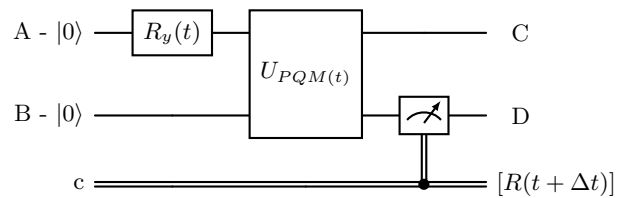
\begin{figure}
\begin{quantikz}[wire types={q,q,c}]
    \lstick{A - $\ket{0}$}&\gate{R_{y}(t)}&\gate[2]{U_{PQM(t)}}&&\rstick{\; C} \\
    \lstick{B - $\ket{0}$}&&&\meter{} \wire[d][1]{c}&\rstick{ \; D} \\
    \lstick{c} &&&\phase{}& \rstick{$[R(t + \Delta t)]$}
    \end{quantikz}
    \caption{Simple circuit representation of a single time-step of the photonic quantum memristor. in Qiskit \cite{ibm-quantum}. The $R_{y}(t)$ gate, rotating the input qubit by a time-dependant amount, is responsible for the input state preparation. The $U_\mathrm{PQM}(t)$ gate performs the two-qubit rotation, encoding the beamsplitter time-dependant reflection and transmission. Finally, a set of measurements is performed on the ancilla qubit, whose outcome is exploited to update the reflectivity. The output is represented by the top qubit, i.e., the state of mode C.}
    \label{fig: circuit}
\end{figure}

The original photonic system described in the previous section has a dynamical evolution, so we look for a protocol to implement a time-dependent process in Qiskit language which, conversely, allows to apply sequences of quantum gates and channels to a set of qubits. The simplest way to achieve that is to divide our time window into small discrete time steps. For each time step, we build a basic two-qubit circuit that has to accomplish the three following tasks, in this order:
\begin{enumerate}
    \item Prepare the initial qubit state into the desired input state given by Eq.~\eqref{eq: input_state_single}.
    \item Using this input qubit and an additional ancilla qubit, execute a two-qubit gate corresponding to the action of the tunable beam splitter with an empty (vacuum) input branch. In our specific single-photon setting, this two-qubit gate can be encoded through a $4\times 4$ unitary matrix that takes into account the exact value of the tunable reflectivity at the time step we are considering (see Sec.~\ref{subsubsec: correspondence} and Eq.~\eqref{eq: memristor_44_matrix} below for details).
    \item Perform a measurement on the ancilla qubit. By storing the outcome of this measurement, we can then update the reflectivity $R(t + \Delta t)$ for the next time step, as explained in the previous section. A basic circuit representation for a single time step is shown in Fig.~\ref{fig: circuit}. With a sufficiently small time step with respect to the considered problem, this method can reproduce the hysteretic behaviour of the PQM. 
\end{enumerate}

\begin{figure}
        \includegraphics[width=0.49\textwidth]{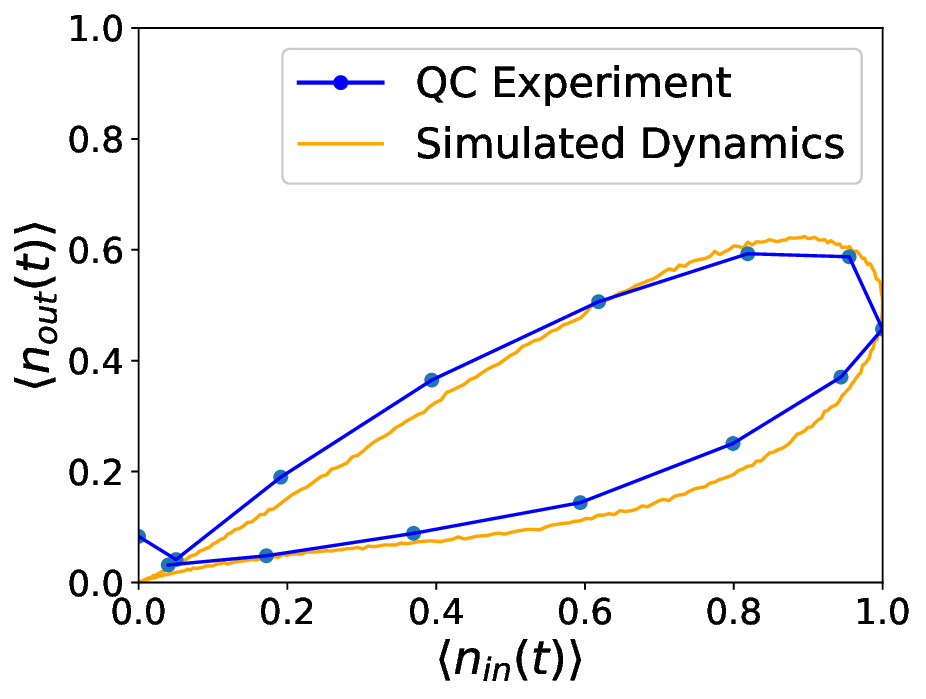}
        \caption{Input-output relationship of the average photon number obtained through a computation on a real quantum computer. The experiment has been executed on May 29th, 2024, using \textit{ibm\_brisbane} with $28$ qubits for the case $T_\mathrm{int}=0.5 T_\mathrm{osc}$ ($T_\mathrm{osc}\equiv 1$). The hysteresis cycle has been divided into $14$ time steps.}
        \label{fig:qiskit_simulation}
\end{figure}

It is known that resetting qubits in order to use them for the next time-step might prove technically difficult in Qiskit \cite{javadi2024quantum}. For this reason, two qubits are required for each time-step of our system's evolution (for example, in the simple case of swapping used qubits with clean ones), meaning that the simulation needs $2n$ qubits to run, where $n$ is the total number of desired steps. In Fig.~\ref{fig:qiskit_simulation} we show our results obtained with an experiment run using \textit{ibm\_brisbane} with $28$ qubits. The hysteretic behaviour is clearly visible, in agreement with the theoretical results. Small deviations can be due the limited amount of time steps considered. In addition, random errors on the execution of quantum operations on the qubits could induce unwanted qubit flips, increasing the error on the statistics collected during the measurement process. Finally, for each time step, the measurement has been repeated $4092$ times, resulting in a lower precision in the reflectivity update.

\subsection{Correspondence between photons and qubits in the PQM}
\label{subsubsec: correspondence}
In this subsection we describe the details of the encoding and the correspondence between qubits and photons in our specific setting.

The main challenge to address is that we have to encode a photonic system (which is bosonic, thus infinite-dimensional) into a gate-based quantum computing architecture, where each qubit is a two level-system. 

As seen earlier, at some fixed time the memristor acts as a beamsplitter with a given reflection amplitude. In the second quantization formalism, this action can be translated into a unitary operator given by \cite{mohan2024digital}
\begin{equation}
    \hat{U}_{\mathrm{B}} = \exp(i \theta(t) (\bd \a + \b \ad)).
\end{equation}
Our objective is to: (i) approximate this exponential operator into a multi-qubit channel and (ii) encode Fock states into two-level systems. 

The problem of translating a $d$-level system into qubits has been already tackled and many encodings have been proposed \cite{sawaya2020resource}. These are usually based on the classical task of writing an integer value (representing the photon number state) into a set of ordered binary numbers (the qubits). 

In our specific system, the observation that the input state is a single-photon state is helpful. Moreover, the mode $B$ of the PQM (i.e., the ancilla) is always in the vacuum state \cite{spagnolo2024quantum}. Thus, for any value of the reflectivity and for any input state in mode $A$ with a photon number equal to one at most, the output of both branches ($C$ and $D$) will also have at most a single photon. 
This fact implies that the evolution of our system remains inside a subspace of photonic states where the maximum number of bosons that can occupy a given mode is exactly one. As a result, we can encode each photon state with a single qubit and use the following unitary rotation matrix to implement the actual evolution of the PQM system in our qubit system:
\begin{equation}
U_\mathrm{PQM} (t) = 
    \begin{pmatrix}
    1 & 0 & 0 & 0\\
    0 & \sqrt{1 - R(t)} & i \sqrt{R(t)} & 0\\
    0 & i \sqrt{R(t)} & \sqrt{1 - R(t)} & 0\\
    0 & 0 & 0 & 1\\
    \end{pmatrix},
    \label{eq: memristor_44_matrix}
\end{equation}
which is expressed in the computational basis $\{\ket{00}, \ket{01}, \ket{10}, \ket{11}\}$, according to the Qiskit standard qubit ordering. 
It is important to note that, since the ancilla qubit starts in state $\ket{0}$, we exclusively use the first and the second column of the above matrix $U_\mathrm{PQM} (t)$, acting on $\ket{00}$ and $\ket{01}$. This is enough to describe the effects of the memristor on the global input state given by the tensor product $\ket{\psi_\mathrm{in}}\otimes\ket{0}$, where $\ket{\psi_\mathrm{in}}$ is the state at mode $A$ given in Eq.~\eqref{eq: input_state_single} and $\ket{0}$ is the vacuum ancilla state in mode $B$. The results of this operation is a global pure state in modes $C$ and $D$, from which the density matrix of the output photon of Eq.~(\ref{outputC}) can be obtained by tracing out the qubit in $D$ (see Fig.~\ref{fig: circuit}).  

Differently, if a photon is present in mode $B$, we would need to take into account Hong-Ou-Mandel (HOM) effect \cite{bouchard2020two}, since we are dealing with indistinguishable photons going through a beamsplitting device (i.e., input states with nonzero probability amplitude for the state $\ket{11}$). This would result in a chance to have two-photon states $\ket{2}_F$ in the same output mode, where the $F$ label refers to Fock number states. To simulate this situation, one can use two qubits to build a one-to-one mapping between qubit states and Fock states with a number of photons ranging from $0$ up to $3$ as \cite{somma2003quantum}
\begin{equation}
    \begin{split}
        & \ket{0}_F \leftrightarrow \ket{00} ,\\
        & \ket{1}_F \leftrightarrow \ket{01} ,\\
        & \ket{2}_F \leftrightarrow \ket{10} ,\\
        & \ket{3}_F \leftrightarrow \ket{11} ,
    \end{split}
\end{equation}
where in the left column we report the photon states and in the right column the corresponding qubit states.
In this work we do not consider such situations, which can be subject of further studies. This choice allows us to reduce the complexity of the problem and keep a one-to-one relationship between photons and qubits. Also, we stress that the first physical implementation of the PQM has been designed with a single optical input \cite{spagnolo2024quantum}.
Nevertheless, interesting results on the digital quantum simulation of a $50/50$ beamsplitter including HOM effect have been reported in Ref.~\cite{mohan2024digital}.

\section{Entanglement dynamics in two distinct photonic quantum memristors}
\label{sec: entangled_section}
In the above sections, in accordance with experimental results, we have seen that the PQM exhibits memristive behaviour when considering two relevant quantities, such as the mean number of input and output photons or the quantum coherence of the input and output states of the photon. In this section we show how a system made of two independent, noninteracting PQMs can manifest memristive behaviour on the response curve of two quantum resources, namely entanglement and quantum coherence, between output and input two-photon states when two photons go through the pair of memristors. In the following, we assume that the effects on the environment on the input photons are negligible (i.e. there is no photon loss or dissipation).

\begin{figure}
         \includegraphics[width=0.40\textwidth]{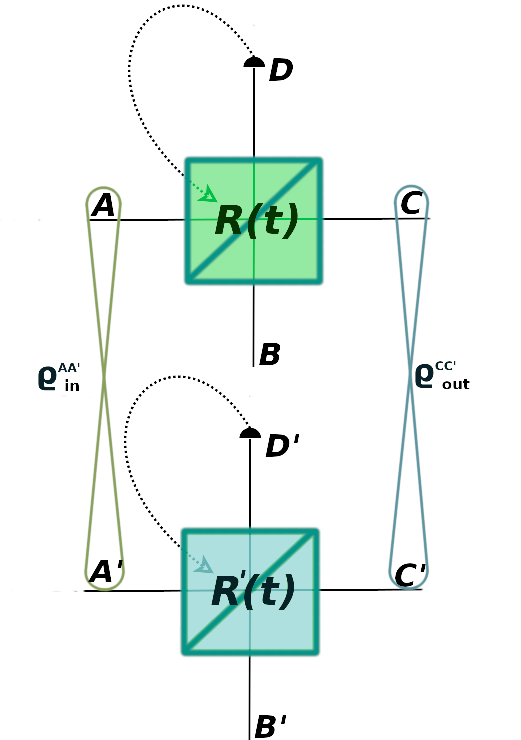}
         \caption{General scheme of the setup used in Sec.~\ref{sec: entangled_section}. The input states going through the two uncoupled PQMs are entangled according to Eq.~\eqref{eq: input_state_double}. Measurements are performed on mode $D$ and $D'$ to update the reflectivities of both PQMs.}
        \label{fig:two_mem}
\end{figure}

Previous works have studied entanglement among two and three quantum memristive subsystems in a superconductive platform \cite{kumar2021entangled, kumar2022tripartite}. Apart from the substantial differences between the considered quantum memristor models, in those works the authors study two (or three) coupled dynamical subsystems where the entanglement is generated due to an interaction between the subsystems themselves. 
In this work, instead, we take an initially entangled pair of photons and investigate the entanglement properties of the output pair in two independent PQMs, that can also be far away from each other. The different framework adopted brings to different physical results, as in the referenced work there is a continuous exchange of information between the two/three parties. This information exchange leads to different phenomena such as shrinking and enlarging of hysteresis cycles and direction changes that we do not observe in the two PQMs assembly.
The system we study is illustrated in Fig.~\ref{fig:two_mem}.

Since the second physical input of each memristor is in the vacuum state (i.e., no incident photons), we can once again encode our photon states into single qubits and, if desired, we can simulate the two-PQM system by adapting and extending the introduced single PQM quantum circuit of Fig.~\ref{fig: circuit}. For this reason, as a paradigmatic example of entanglement between particles, we choose the four Bell states as our input states, that is
\begin{equation}
    \begin{split}
            & \ket{\Psi^{\pm}_{\mathrm{in}}} = \alpha (t) \ket{01} \pm \beta (t) \ket{10}, \\
            & \ket{\Phi^{\pm}_{\mathrm{in}}} = \alpha (t) \ket{00} \pm \beta (t) \ket{11}. 
    \end{split}
    \label{eq: input_state_double}
\end{equation} 
By selecting oscillating functions for $\alpha(t)$ and $\beta(t)$, we can have at the same time two important features: (i) oscillating average photon number for each of the input modes and (ii) periodic entanglement and coherence between the two photons. Specifically, we set
\begin{equation}
       \alpha (t) = \sin{ \left( \frac{\pi t}{T_{\mathrm{osc}} } \right) }, \quad
       \beta (t) = \cos{ \left( \frac{\pi t}{T_{\mathrm{osc}} }\right)}.
\end{equation}
With this specific choice, the average photon number oscillates with the usual period $T_\mathrm{osc}$. Moreover, the input state oscillates between a completely disentagled state and a maximally entangled one. 

To explicitly quantify this process, we adopt concurrence as a measure of bipartite entanglement \cite{nielsen2010quantum, horodecki2009quantum}. Since we restrict our interest to two-qubit states, encoding two single photon Fock states, a generic entangled state can be expressed as
\begin{equation}
    \ket{\psi} = a \ket{00} + b\ket{01} + c \ket{10} + d \ket{11}.
\end{equation}
In such cases the concurrence is simply
\begin{equation}
    C ( \ket{\psi}) = 2 \abs{ad - bc},
\end{equation}
which, for both states of Eq.~(\ref{eq: input_state_double}), gives the input concurrence 
\begin{equation}
    C_{\mathrm{in}} (t) = 2 \abs{\alpha(t)\beta(t)} = 2 \abs{\cos{ \left( \frac{t}{T_{\mathrm{osc}} }\pi \right) } \sin{ \left( \frac{t}{T_{\mathrm{osc}} }\pi \right)}}.
\end{equation}
The input concurrence in this case is exactly identical to the input coherence, computed through the $\mathcal{C}_{l_1, in}=\sum_{i\neq j}|\rho_{\mathrm{in},ij}|$ summing the absolute values of the off-diagonal elements of the two-qubit initial density matrix  \cite{baumgratz2014quantifying,streltsov2017colloquium}.

Due to the structure of $C_\mathrm{in}(t)$ and $\mathcal{C}_{l_1,\mathrm{in}}(t)$, the period of entanglement and coherence oscillations is $T_\mathrm{osc}/2$. This means that for one full oscillation of the input state, the concurrence and the $l_1$-\textit{norm} of coherence will oscillate twice between their minimum ($0$) and maximum ($1$) value.
To evaluate the concurrence of the output state (i.e., after that both photons have gone through the PQMs), we determine the output density matrix of the relevant part of the global system. This is done by applying the beamsplitting process on the full input system (i.e., both the input photons and the ancilla states) and then tracing out over the measured modes degrees of freedom. In the language of quantum channels, we can express the output density matrix as
\begin{equation}
    \rho_{\mathrm{out, C, C'}} = \mathrm{Tr}_{D, D'} \{ \hat{U} (t) \hat{U}' (t) \rho_{\mathrm{in, A, A'}} \hat{U}{\dagger} (t) \hat{U}'^{\dagger}(t)\},
\end{equation}
where $\hat{U}$ and $\hat{U}'$ are the unitary matrices representing the single PQM beamsplitting process at some value of the reflectivities $R(t)$ at $R'(t)$. The two unitary matrices commute (since the PQMs are not interacting) and the feedback is obtained through a classical communication. This implies that the whole evolution can be described as a multipartite LOCC process, which could be extended to a multiple PQMs scenario.

\begin{figure}
         \includegraphics[width=0.48\textwidth]{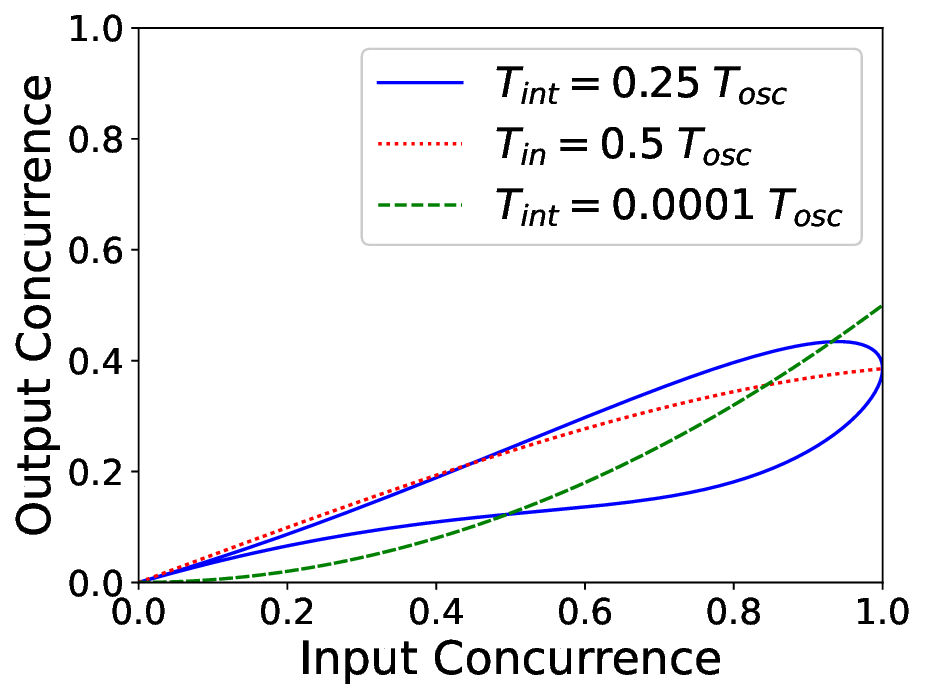}
         \includegraphics[width=0.48\textwidth]{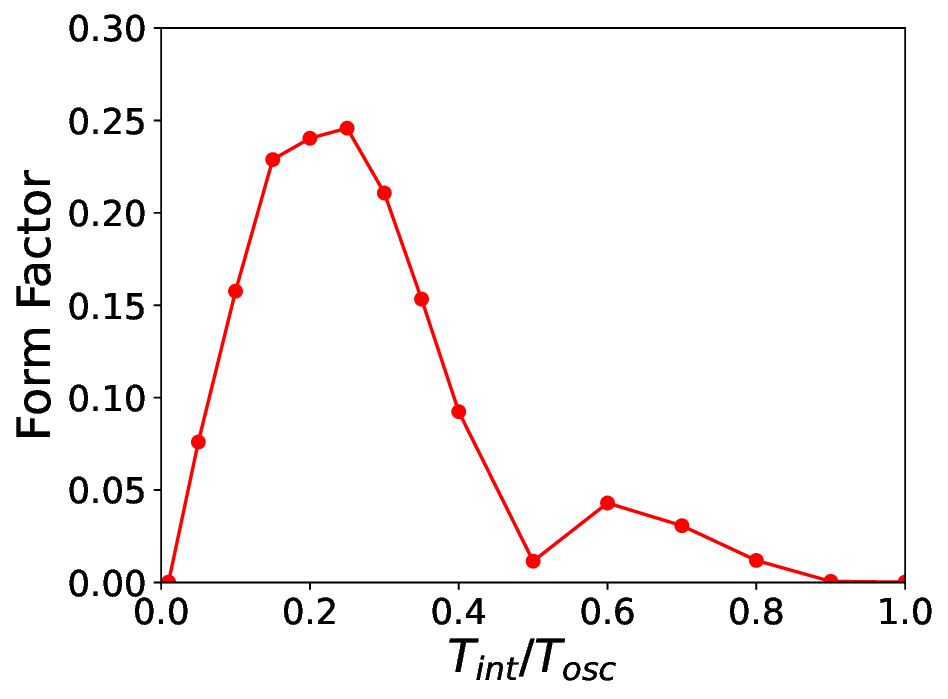}
         \caption{Upper panel: Input-output relations for a couple of entangled photons going through two distinct PQMs. The behaviour changes for different choices of the internal memory time of the memristors. The blue line corresponds to the maximum form factor for this setup. The red dotted line represents the low period regime, while the green dashed line is obtained in the low period regime. No difference in the dynamics has been detected between $\ket{\Psi^{+}_\mathrm{in}}$ and $\ket{\Psi^{-}_\mathrm{in}}$. Lower panel: Form factor for the concurrence and coherence hysteresis cycles related to the state $\ket{\Psi^{\pm}_{in}}$. There is a maximum for $T_\mathrm{int} = 0.25\;T_\mathrm{osc}$. The results shown in these plots are obtained through simulations based both on the results shown in Eq.~\eqref{eq: analytical_concurrence} and simulation on Qiskit}
        \label{fig:concurrence_01_10}
\end{figure}

\begin{figure}
         \includegraphics[width=0.48\textwidth]{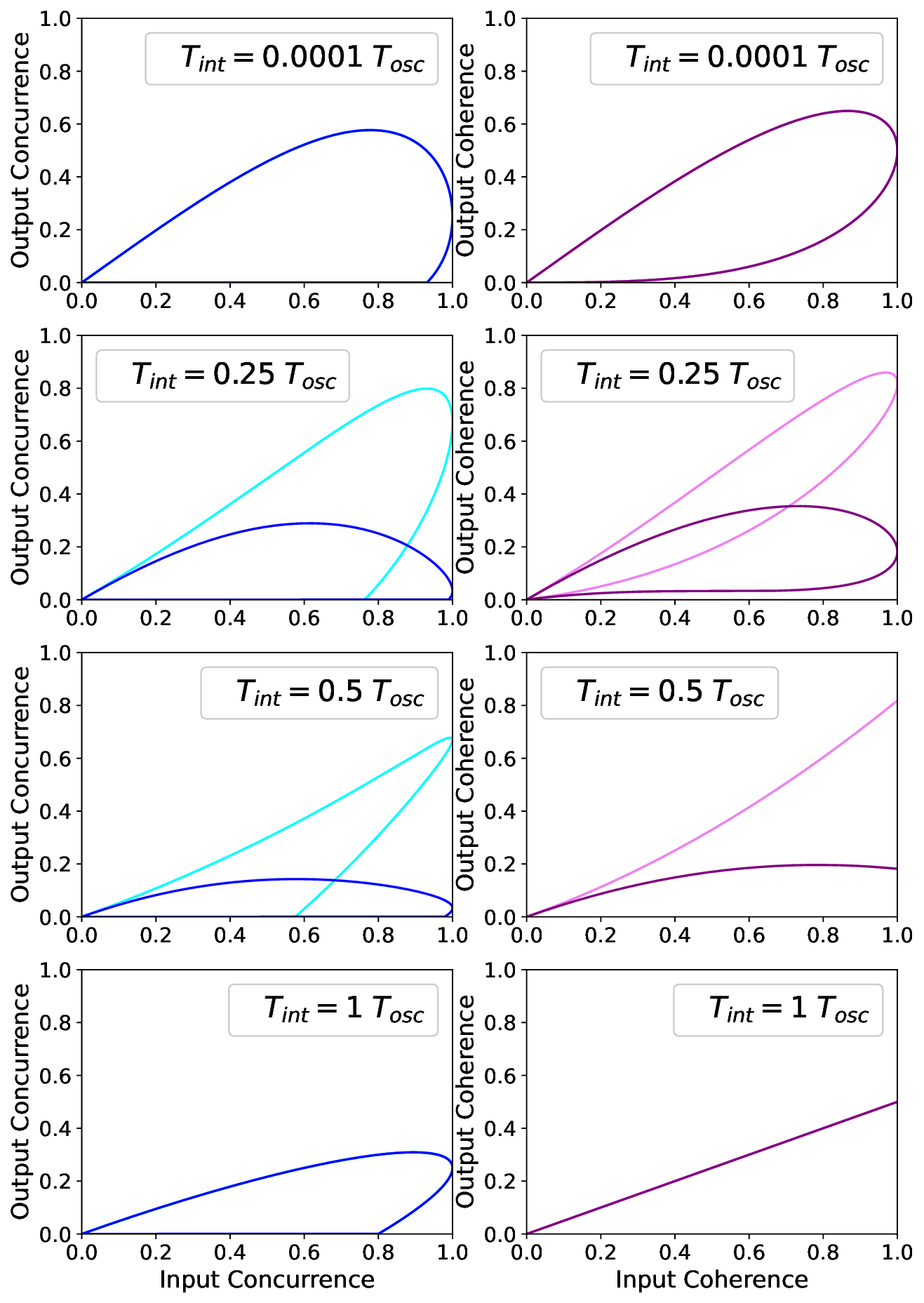}
         \includegraphics[width=0.48\textwidth]{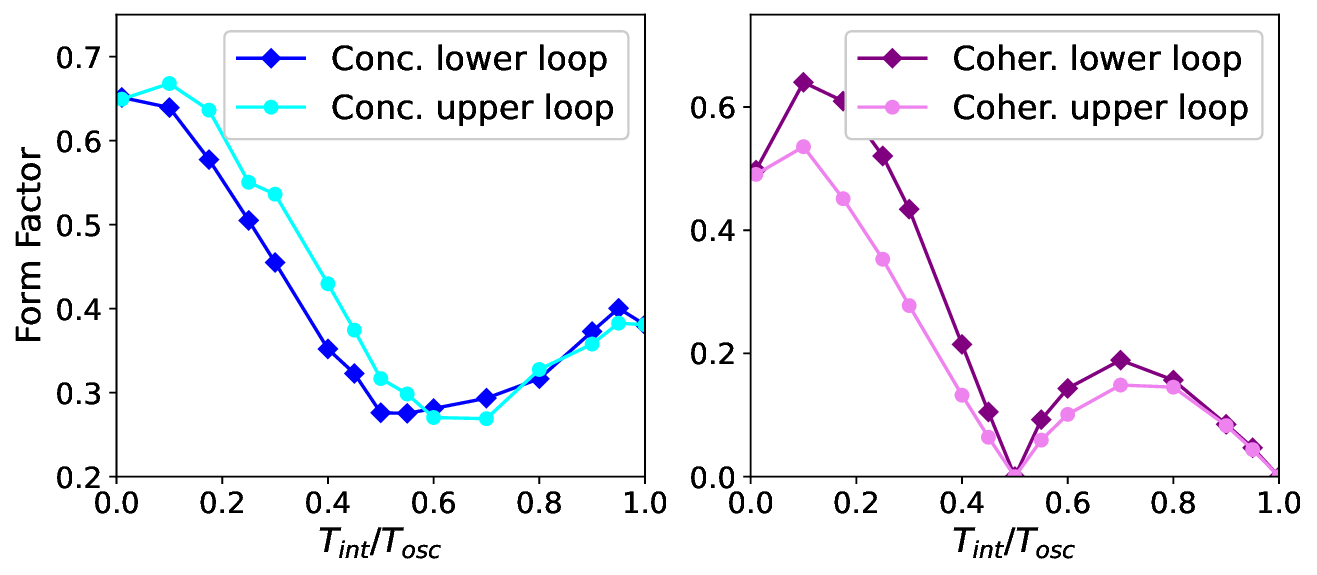}
         \caption{Upper block: Input-output relations between the concurrence (left column, blue lines) and the coherence (right column, purple lines) at different fractions of $T_\mathrm{int}/T_\mathrm{osc}$ for the $\ket{\Phi^{\pm}_\mathrm{in}}$ input state. No difference in the dynamics due to relative phase has been detected. Lower block: Form factor for the concurrence and coherence hysteresis cycles related to the state $\ket{\Phi^{\pm}_\mathrm{in}}$. The dynamical regimes are described in more detail in Sec. \ref{sec: entangled_section}. The results shown in these plots are obtained through simulations based both on the results shown in Eq.~\eqref{eq: analytical_concurrence} and through simulation on Qiskit} 
        \label{fig:concurrence_00_11}. 
\end{figure}

Having the output density matrix, the output concurrence can be computed as \cite{horodecki2009quantum, hill1997entanglement}
\begin{equation}
    C_\mathrm{out} (t) = \max\left\{0,2\lambda_\mathrm{max} - \text{Tr}(M) \right\},
\end{equation}
where $M = \sqrt{\sqrt{\rho_\mathrm{out}}\Tilde{\rho}_\mathrm{out}\sqrt{\rho_\mathrm{out}}}$. The greatest eigenvalue of $M$ is denoted as $\lambda_\mathrm{max}$, while $\Tilde{\rho_\mathrm{out}} = (\sigma_y \otimes \sigma_y)\rho_\mathrm{out}^{*}(\sigma_y \otimes \sigma_y)$. According to this definition, it is clear that $C_{out}$ is a non-negative quantity with $0 \leq C_\mathrm{out}\leq 1$. The output coherence is calculated via the usual $l_1$-\textit{norm} $\mathcal{C}_{l_1,\mathrm{out}}(t)$ for the two-qubit output state. With the sole assumption that $\alpha$ and $\beta$ are real numbers, this leads us to 
\begin{equation}
    \begin{split}
        & C_\mathrm{out, \ket{\Psi_{\pm}}} (t) = 2 \abs{\alpha(t)}\abs{\beta(t)} \sqrt{1 - R(t)} \sqrt{1 - R'(t)},\\
        & C_\mathrm{out, \ket{\Phi_{\pm}}} (t) = 2 \abs{\beta(t)}\sqrt{1 - R(t)} \sqrt{1 - R'(t)}\times \\
        & \quad\quad\quad \times(\abs{\alpha(t)} - \abs{\beta(t)}\sqrt{R(t)} \sqrt{R'(t)}),
        \label{eq: analytical_concurrence}
    \end{split}
\end{equation}
while, for the coherence, we get the same expression in both cases
\begin{equation}
    \mathcal{C}_{l_{1}}(\rho_{\mathrm{out, C,C'}}) = 2 \abs{\alpha(t)\beta(t)}\sqrt{1 - R(t)}\sqrt{1 - R'(t)}.
\end{equation}
It is important to remember that the explicit expression of $R(t)$ and $R'(t)$ is different according to the input state. The results obtained remain general even for Reflectivity update rules different from the one shown in Eq.~(\ref{eq: reflectivity_equation})
This shows us that memristive behaviour can be manifested also at the two memristor level in the case of input Bell states for the concurrence and the coherence.
Numerical results of our analysis for the input states shown in Eq.~(\ref{eq: input_state_double}) are presented in Figs.~\ref{fig:concurrence_01_10} and  \ref{fig:concurrence_00_11}. 

In order to evaluate the potential memory content of the resulting hysteresis cycles, we adopt the form factor \cite{biolek2014interpreting, fomichev1997equation}
\begin{equation}
    \mathcal{F} = 4 \pi \frac{A}{P^2},
\end{equation}
which is a measure of how large the area $A$ inside the hysteresis cycle is with respect to its perimeter $P$. The same approach has been used to study different models of quantum memristors \cite{kumar2021entangled, kumar2022tripartite}. Here, we adopt $\mathcal{F}$ for the concurrence and coherence hysteresis cycles. Specifically, the form factor provides an easy tool for tuning the relative periods ($T_\mathrm{int}$ and $T_\mathrm{osc}$) to maximize the memory content.

We start considering the results obtained for the $\ket{\Psi^{\pm}_\mathrm{in}}$ states, shown in Fig.~\ref{fig:concurrence_01_10}. The behaviour of the input-output relation of the concurrence is similar to the average photon number in the single memristor scenario (central panel of Fig.~\ref{fig:sumup_single}). We can distinguish a high period regime (when $T_\mathrm{int} \gtrsim 0.5 \; T_\mathrm{osc}$) when the behaviour of the memristor tends to be similar to a $50/50$ beamsplitter (i.e. with no memory effects). This can be attributed to the fact that we are sending a periodic input such that both reflectivities become constant as a result of integrating over one or more than one concurrence oscillation period. For lower values of $T_\mathrm{int}$ a pinched hysteresis loop opens up, reaching his peak form factor at $T_\mathrm{int} = 0.25 \; T_\mathrm{osc}$.
For very small periods ($T_\mathrm{int} \lesssim 10^{-2} \; T_\mathrm{osc}$), the two PQMs lose memory effects since the loop closes, resulting in a liner relationship. 

Moving on to the characterization of the $\ket{\Phi_\mathrm{in}^{\pm}}$ input state, displayed in Fig.~\ref{fig:concurrence_00_11}, we observe that the dynamics is richer and we can distinguish two different loops when looking at the concurrence and coherence dynamics over $T_\mathrm{osc}$. This is due to the fact that, for our specific choice of $\alpha(t)$ and $\beta(t)$, the two quantum resources oscillate with half the period with respect to the average photon numbers, giving rise to two shorter cycles. Note that the $\ket{\Psi_\mathrm{in}^{\pm}}$ state is perfectly symmetrical under this point of view, since $\ket{01}$ and $\ket{10}$ produce the same effects, respectively, on the first and second memristor and thus the two cycles are perfectly identical. The same cannot be said for $\ket{\Phi_\mathrm{in}^{\pm}}$, since its components $\ket{00}$ and $\ket{11}$ have a different photon number. 
Starting with the concurrence input-output relation (left column of Fig.~\ref{fig:concurrence_00_11}), we see that in the high period range ($T_\mathrm{int} \gtrsim T_\mathrm{osc}$, i.e., the 50/50 beamsplitter scenario) the two loops overlap. Passing to lower values of $T_\mathrm{int}$ the cycles start separating, until they overlap again with a larger form factor for low periods (bottom left panel of Fig.~\ref{fig:concurrence_00_11}). 
Instead, the coherence response curve (right column of Fig.~\ref{fig:concurrence_00_11}) follows a linear behaviour for high $T_\mathrm{int}$. By lowering this memory interval of our PQMs, we witness two progressively larger hysteresis loops that close into two lines when $T_\mathrm{int} \sim 0.5 \; T_\mathrm{osc}$. For even smaller $T_\mathrm{int}$ they open up again, finally overlapping in the low period regime. For both quantum resources, we find that the form factor tends to be higher for lower memory periods (bottom right panel of Fig.~\ref{fig:concurrence_00_11}). This behaviour is similar to what happens for the coherence in the single PQM (see Fig.~\ref{fig:coherence_single}).

\section{Conclusions}
\label{sec: conclusions}
In this work we have shown that a single photonic quantum memristor (PQM) manifests hysteretical dynamics on the coherence, which is a purely quantum property.
Subsequently, we have seen that the entanglement between two distinct photons going through two different and uncorrelated PQMs has a non-Markovian dynamics as well.
We have adopted the form factor as a quantification tool for the memory capacity of hysteresis cycles. In our specific situations we show that the memory effects are extended to strictly quantum resources, such as entanglement and coherence, underlining the potential of the PQM-based architecture for processing quantum information. 

We have also presented a circuit model capable of simulating the dynamics of one or more PQMs and tested it on a real quantum computer.
The interest in our quantum circuit conversion of the PQM is twofold: On the one hand, it is a good example of how quantum computers can successfully simulate the dynamics of a quantum system even in an open scenario. On the other hand, it points out how studying open quantum systems can enrich the landscape of processes that quantum computers can perform, adding memory effects and hysteretical phenomena to an already impressing list that grows day by day thanks to a wide research field.
Other works \cite{li2021simulation, guo2022quantum} have proposed quantum circuit models capable of manifesting memristive properties in the average values of some observables. We remark that the starting point of our study is to thoroughly characterize a physical, feasible, and scalable memristive system which is different from the ones considered in previous works: namely, a photonic circuitry platform which manipulates quantum states of traveling photons \cite{spagnolo2022experimental}.

Overall, in this work we have limited ourselves to relatively simple settings, since we fixed the function describing the time evolution of the reflectivity and the input's probability amplitudes. 
Evaluating the PQM response in different settings and with different approaches might give us insight on which tasks can become more efficient through its implementation and this, in turn, can provide more information on how to exploit the memory effects emerging from the PQM's dynamics.

Additionally, to better evaluate the potential of the PQM as the building block for neuromorphic quantum computing, the scalability of this system must be considered. Given the promising results of the simple uncoupled scenario, interesting behaviour could emerge when considering a network of connected PQMs.

Finally, our results pave the way to further analyses about the exploitation of scalable PQM-based networks for scopes of quantum machine learning and quantum neural networks \cite{hernani2024machine}. As an example, classical memristors have been proposed as candidates for various tasks, such as Hopfield Network components \cite{li2022tristable, cai2020power}. Further efforts could be made in the direction of a quantum Hopfield Network based on Quantum Memristors. Additionally, due to its nonlinear dynamics and memory capacity, the PQM has already been used in the context of Reservoir Quantum Computing \cite{spagnolo2022experimental}. Further work could assess the impact of entanglement between input states as a computational resource and the specific features of the PQM (the update rule for the Reflectivity and/or the feedback itself) could be tuned for optimized efficiency in specific recognition or classification tasks.

\begin{acknowledgements}
The code used in this manuscript is available upon request. 
R.L.F. acknowledges support by MUR (Ministero dell’Università e della Ricerca) through the following projects: PNRR Project ICON-Q – Partenariato Esteso NQSTI – PE00000023 – Spoke 2, PNRR Project AQuSDIT – Partenariato Esteso SERICS – PE00000014 – Spoke 5, PNRR Project PRISM – Partenariato Esteso RESTART – PE00000001 – Spoke 4. A.F. aknowledges useful discussions with Michał Siemaszko.
\end{acknowledgements}

\providecommand{\noopsort}[1]{}\providecommand{\singleletter}[1]{#1}%

\end{document}